\documentstyle[12pt]{article}
\begin{document}
\renewcommand{\thefootnote}{\fnsymbol{footnote}}
\def\lambdabar{{\mathchar'26\mkern-9mu\lambda}}
\begin{center}ON THE SPIN-ROTATION-GRAVITY COUPLING\footnote{This
paper is based on a lecture delivered at the Mexican Meeting on Gauge
Theories of Gravity (Mexico City, October 6--10, 1997).}\\

\vspace{.5in}

Bahram Mashhoon\\ Department of Physics and Astronomy\\ University of Missouri-Columbia\\
Columbia, Missouri 65211, USA\\
\vspace{1in}

ABSTRACT\\
 \end{center}

The inertial and gravitational properties of intrinsic spin are discussed and some of the recent
work in this area is briefly reviewed.  The extension of relativistic wave equations to
accelerated systems and gravitational fields is critically examined.  A nonlocal theory of
accelerated observers is presented and its predictions are compared with observation.\\

\newpage

The inertial and gravitational couplings of intrinsic spin have recently received attention as
experimental evidence for spin-rotation coupling has become available.  Friedrich W.\ Hehl has
made significant contributions to this important topic and it is therefore a great pleasure for
me to dedicate this paper to him on the occasion of his sixtieth birthday.

\vspace{12pt}

\noindent 1. INTRODUCTION
\vspace{10pt}

The spin-rotation-gravity coupling has appeared in the work of many authors who have been mainly
interested in the study of wave equations in accelerated systems and gravitational fields [1]. 
Indeed, the coupling under consideration here directly involves wave effects that pertain to the
physical foundations of general relativity.  Classically, motion occurs via particles as well as
electromagnetic waves.  The basic geometric structure of Einstein's theory of gravitation
accords a special status to the motion of classical test particles and null rays, since these
idealized physical systems follow geodesic paths that are intrinsic to the geometry of the
spacetime manifold [2].  In contrast, the motion of a wave packet in general relativity does not
pertain to intrinsic geometric properties of the spacetime.  Can one provide a purely geometric
description of {\it diffraction} phenomena, for instance?  To illustrate the problem, let us
consider the following thought experiment:  Imagine a ray of light that has frequency $\omega$
according to observer {\it O} and the class of observers boosted with respect to {\it O} at the
same event along the direction of propagation of the ray.  The frequency measured by any such
observer is $\omega^\prime=\gamma\omega(1-\beta)$ in accordance with the Doppler effect.  It
follows that the wavelength of the radiation can become extremely large or extremely small
according to the boosted observers; however, the respective limiting values of infinity and zero
are excluded since $|\beta|<1$.  On the other hand, it can be shown that the effective radius of
curvature of spacetime as measured by the boosted observers is generally Lorentz contracted [3]. 
According to all observers, however, the worldline of the ray is a null geodesic even when the
measured wavelength far exceeds the measured radius of curvature.  The only physical conclusion
that one can draw from this analysis is that the wavelength of the radiation must be zero for all
observers in order that the complete absence of diffraction can be satisfactorily explained. 
Thus null geodesics would carry infinite energy in the quantum theory; hence, the standard
axiomatic formulations of general relativity in terms of clocks and light rays are physically
unrealistic.

Einstein formulated general relativity as a theory of pointlike coincidences [2]; therefore, the
theory is most consistent when wave phenomena, which generally require extended intervals of
space and time for their characterization, are treated in the eikonal limit.  In general, wave
phenomena in a gravitational field depend upon the observer; moreover, a completely covariant
analysis is not possible since an observer can set up an admissible coordinate system in its
neighborhood only within a spatial region of radius $R\ll\cal L$, where $\cal L$ is an
acceleration length, and only wavelengths $\lambda<R$ can then be determined by the observer.

Consider, for the sake of simplicity and the exclusion of matter-related effects, the scattering
of electromagnetic radiation from a black hole in terms of the standard set of inertial
observers in the asymptotically flat region of the spacetime.  It turns out that for a
Schwarzschild black hole the amplitudes for the scattering of right circularly polarized (RCP)
and left circularly polarized (LCP) waves are equal and hence the spherical symmetry of this
field preserves the polarization of the incident radiation in the scattered waves.  However, for
a Kerr black hole RCP and LCP radiations are scattered differently.  This can be traced back to
the influence of a gravitational coupling between the intrinsic spin of the radiation field and
the rotation of the source.  In this way, the deflection of the radiation by a rotating mass
becomes polarization dependent [4].  Imagine a rotating body with mass $M$ and angular momentum
${\bf J} =  J{\bf \hat z}$ with its center of mass at the origin of
coordinates and a beam of radiation propagating above the body nearly parallel to the x-axis with
impact parameter $D$.  The Einstein deflection angle for the beam is $\Delta = 4GM/c^2D$; however,
RCP radiation is essentially deflected by an angle $\Delta - \delta$ and LCP radiation by $\Delta
+ \delta$, where $\delta = 4\lambdabar GJ/c^3D^3$.  In the JWKB limit,
$\delta\rightarrow 0$ and the principle of equivalence is recovered.  The differential
deflection of polarized radiation is very small; e.g., it is of order one milliarcsecond for
radio waves with $\lambdabar\sim 1$ cm passing just over the poles of a rapidly rotating neutron
star.  Upper limits on the deviation from the principle of equivalence for polarized radio
waves deflected by the Sun have been placed by Harwit {\it et al}. [5].  Astrophysical
implications of this effect have been considered by a number of authors [6]; in particular, it
may become interesting in connection with microlensing with polarized radiation [7].

The differential deflection of polarized radiation is a consequence of the coupling of photon
helicity with the gravitomagnetic field of a rotating mass ${\bf B}_g =
c\mbox{\boldmath $\Omega$}_P$, where 

\begin{eqnarray}
\mbox{\boldmath $\Omega$}_{P}={\frac{GJ}{c^2r^3}}\left [3({\bf \hat
J}\cdot{\bf \hat r}){\bf \hat r} -{\bf \hat J}\right ]
\end{eqnarray}

\noindent is the precession frequency of a free test gyroscope at position {\bf r}. 
According to the gravitational Larmor theorem [4], a gravitomagnetic field can be locally
replaced by a frame rotating at frequency $\mbox{\boldmath$\Omega$}_L =
-\mbox{\boldmath$\Omega$}_P$.  It follows that similar spin-rotation coupling effects are expected
in a rotating frame of reference.  This may be illustrated with a thought experiment:  Consider an
inertial reference frame $\cal S$ and an observer rotating in the positive sense about the
direction of propagation of a plane monochromatic electromagnetic wave of frequency $\omega$.  We
are interested in the frequency of the radiation as measured by the rotating observer.  Special
relativity is based on Poincar\'{e} invariance and the {\it hypothesis of locality}.  The latter
states that an accelerated observer in Minkowski spacetime is at each event equivalent to a
momentarily comoving inertial observer.  Thus the rotating observer is instantaneously inertial
and the transformation between this local inertial frame $\cal S^\prime$ and $\cal S$ results in
the transverse Doppler formula, $\omega^{\prime}=\gamma\omega$, for the frequency of the
radiation.  On the other hand, the observer needs to measure at least several oscillations of the
wave before an estimate for
$\omega^{\prime}$ could be computed from the data.  It follows from this line of argument that
the transverse Doppler formula must be valid in the eikonal limit.  It is more reasonable to
assume that the hypothesis of locality applies to the field at each event; then, 

\begin{eqnarray} F_{(\alpha)(\beta)}(\tau) =
F_{\mu\nu}\lambda^{\mu}_{(\alpha)}\lambda^{\nu}_{(\beta)}\;\; ,
\end{eqnarray}

\noindent which is the projection of the Faraday tensor on the tetrad frame of the
observer, is Fourier analyzed over the proper time $\tau$ of the accelerated observer to
determine its frequency content.  This is the extended hypothesis of locality for wave phenomena
and provides the physical basis for the extension of relativistic wave equations to accelerated
frames and gravitational fields (``minimal coupling'').  For the thought experiment under
consideration, we find in this way that $\omega^{\prime}=\gamma(\omega\mp\Omega)$, where the
upper (lower) sign refers to RCP (LCP) incident radiation.  This result has a simple physical
interpretation:  The electric and magnetic fields rotate in the positive sense with frequency
$\omega$ about the direction of propagation in a plane RCP wave; therefore, from the viewpoint
of the rotating observer the radiation is also RCP but with frequency
$\omega^{\prime}=\gamma(\omega-\Omega)$.  Here the Lorentz factor takes due account of time
dilation.  A similar argument for the LCP radiation leads to the addition of frequencies and
$\omega^{\prime}=\gamma(\omega + \Omega)$.  In terms of the photon energy
$E^{\prime}=\gamma(E\mp\hbar\Omega)$, so that the helicity of the radiation couples to rotation
producing an effect that goes beyond the eikonal limit.  That is
$\omega^{\prime}=\gamma\omega(1\mp\lambdabar/\cal L)$, where ${\cal L}=c/\Omega$ is the
acceleration length of the observer.  It is important to point out that experimental evidence
for such wave effects due to helicity-rotation coupling with $\lambdabar\ll\cal L$ already exists
for microwaves as well as light and will be described elsewhere [8].

It is possible to show that for an arbitrary direction of incidence

\begin{eqnarray}
\omega^{\prime}=\gamma(\omega - m\Omega)\;\;,
\end{eqnarray}

\noindent where $m$ is a parameter characterizing the component of the total angular momentum of
the radiation field along the direction of rotation (``magnetic quantum number'').  For a scalar
or a vector field, $m=0, \pm 1, \pm 2, ...$, while for a Dirac field $m\mp {1\over 2}= 0, \pm 1,
\pm 2,...\;$.  Thus $\omega^\prime$ could be negative, zero or positive.  In the case of a
linearized gravitational radiation field, the helicity-rotation coupling has interesting
consequences for celestial mechanics [9].

The observational consequences of spin-rotation coupling for neutron interferometry in a
rotating frame of reference have been explored in connection with the assumptions that underlie
the physical interpretation of wave equations in an arbitrary frame of reference [10].  In
general, the spin-rotation phase shift is smaller than the Sagnac shift [11] by roughly the
ratio of the wavelength to the dimension of the interferometer.

A proper theoretical treatment of the inertial properties of a Dirac particle is due to Hehl
and his collaborators [12].  This treatment has been extended in several important directions
by a number of investigators [13-16].  The significance of spin-rotation coupling for atomic
physics has been pointed out by Silverman [17].  Moreover, the astrophysical consequences of the
helicity flip of massive neutrinos as a consequence of spin-rotation coupling have been
investigated by Papini and his collaborators [18].  Bell and Leinaas [19] attempted to explain
certain depolarization phenomena in circular accelerators in terms of a thermal bath caused by
the centripetal acceleration of the (polarized) particles involved; however, Papini {\it et al.}
[20] have shown that the data should be interpreted instead in favor of spin-rotation coupling. 
In fact, there is no experimental evidence for an acceleration-induced thermal ambience at
present; moreover, it does not come about in the theoretical structures discussed in this paper. 
To appreciate this point, imagine the energy-momentum tensor of the field as measured by an
accelerated observer $T_{(\alpha)(\beta)}=T_{\mu\nu}\lambda^{\mu}_{(\alpha)}\lambda^{\nu}_{(\beta)}$;
once the field is absent in the inertial frame, the energy-momentum measured by any standard device
vanishes.  A similar result involving the vacuum expectation value of the energy-momentum tensor is
expected to hold in the quantum theory. 

Direct evidence for the coupling of intrinsic spin to the rotation of the Earth has recently
become available [21, 22].  In fact, according to the natural extension of general relativity
under consideration here, every spin-$1\over 2$ particle in the laboratory has an additional
interaction Hamiltonian

\begin{eqnarray} H\simeq
-\mbox{\boldmath $\sigma$}\cdot\mbox{\boldmath $\Omega$}_{\oplus}+\mbox{\boldmath
$\sigma$}\cdot\mbox{\boldmath $\Omega$}_{P}\;\;,
\end{eqnarray}

\noindent where $\hbar\Omega_{\oplus}\sim 10^{-19}$eV and $\hbar\Omega_{P}\sim 10^{-29}$eV for
the gravitomagnetic field of the Earth.  The observation of the extremely small gravitomagnetic
Stern-Gerlach force $-\nabla(\mbox{\boldmath $\sigma$}\cdot\mbox{\boldmath $\Omega$}_{P})$ would
be of basic interest since it would demonstrate that the spin part of the gravitational
acceleration is not universal: particles in different spin states fall differently in the
gravitational field of the Earth.  This quantum gravitational force has a classical analog in the
Mathisson-Papapetrou force.
\vspace{12pt}

\noindent 2. CAN LIGHT STAND STILL?
\vspace{10pt}

An important consequence of the general formula (3) for $\omega^\prime$ is that $\omega^\prime$
can be negative or zero.  Since rotation is absolute and there is therefore an absolute
distinction between the rotating observers and the inertial observers, negative $\omega^\prime$
cannot be excluded.  A comment is in order here regarding the formal possibility of
reinterpreting radiation with negative $\omega^\prime$ as positive frequency radiation
propagating in the opposite direction.  This would imply that the causal sequence of events
would depend upon the motion of the observer; moreover, to keep $\omega^\prime$ positive in all
cases one has to assume that the observer-dependent causal sequence is also dependent upon the
details of the physical process under consideration.  This is hardly acceptable physically and
it appears more consistent to simply admit to the possibility of existence of negative energy
states according to {\it noninertial} observers.

Let us next consider the possibility that $\omega^{\prime} =0$ for $\omega = m\Omega$ in equation (3);
that is, the radiation can stand still for a rotating observer.  For instance, in the thought
experiment involving the uniformly rotating observer, a positive helicity wave of frequency $\omega =
\Omega$ would stand completely still due to a mere rotation of the observer.  There is no
experimental evidence in support of this circumstance.

It is possible to interpret the classical theory of Lorentz invariance in terms of the relative
motion of the inertial particles and the absolute motion of electromagnetic waves.  The motion
of radiation is absolute in the sense that it is independent of any inertial observer.  This
basic consequence of Lorentz invariance can be generalized to all observers and raised to the
status of a physical principle that would then exclude the possibility that a fundamental
radiation field could stand completely still with respect to an accelerated observer [23].  It
is important to describe briefly how such a physical principle would fit in with the foundations
of the theory of relativity.  The idea of {\it relativity} has to do with the
possibility of changing one's standpoint for the purpose of observation.  This is kinematically
permissible with classical point particles, since an observer can stay at rest with a classical
particle.  In fact, Minkowski elevated this circumstance to the status of an axiom [24].  On the
other hand, Lorentz invariance implies that an inertial observer can never stay at rest with
respect to a classical electromagnetic wave.  In this sense, the motion of the wave is
nonrelative, i.e. {\it absolute}.  These issues are related to an important observation
due to Mach [25]:  The intrinsic state of a Newtonian point particle, i.e. its mass, is not
directly related to its extrinsic state $({\bf x}, {\bf v})$ in absolute space and
time.  Let us note that this extrinsic state could therefore be shared by any observer, say,
that would momentarily stay at rest with the particle.  Extending Mach's observation to the
case of an electromagnetic wave, we note that the intrinsic properties of a wave, i.e. its
frequency, wavelength, amplitude and polarization, are directly related to its extrinsic state
in (absolute) time and space $\psi(t,{\bf x})$.  Our basic assumption then implies that
this state of the wave cannot be ``shared'' by a local observer in the sense that regardless of
its motion the observer can never stay at rest with the electromagnetic wave.  The duality of
classical particles and waves can thus be extended to their motion as well and our basic
postulate may be stated in terms of the principle of complementarity of absolute and relative
motion [23].  

To implement this physical principle, it is necessary to take a more general view of the
relationship between accelerated and inertial observers.  The basic laws of physics have been
formulated with respect to inertial systems; therefore, accelerated observers must be linked to
inertial observers and the hypothesis of locality provides the first step in this process.  A
more general treatment leads to the nonlocal theory of accelerated observers.

\vspace{12pt}

\noindent 3. ACCELERATED OBSERVERS AND NONLOCALITY
\vspace{10pt}

Let us suppose that a pulse of electromagnetic radiation is incident on an accelerated observer
in Minkowski spacetime.  The observer determines the field amplitude to be ${\cal
F}_{\alpha\beta}(\tau)$.  Let $F^{\prime}_{\alpha\beta}(\tau)=F_{(\alpha)(\beta)}(\tau)$ be the
field amplitude instantaneously measured by the momentarily comoving inertial observers.  The
accelerated observer passes through a continuous infinity of momentarily comoving inertial
observers; therefore, the most general linear relationship between $\cal F_{\alpha\beta}$ and
$F^{\prime}_{\alpha\beta}$ consistent with causality is

\begin{eqnarray}   
{\cal F}_{\alpha\beta}(\tau)=F^{\prime}_{\alpha\beta}(\tau)+\int^{\tau}_{\tau_{0}}{\cal
K}_{\alpha\beta}^{\;\;\;\;\:\gamma\delta}(\tau,\tau^{\prime})F^{\prime}_{\gamma\delta}(\tau^{\prime})d\tau^{\prime}\;\;,
\end{eqnarray}

\noindent where $\tau_0$ is the initial instant of accelerated motion.  It is expected that the
kernel $\cal K$ would be directly related to the acceleration of the observer and so the
nonlocal part would in general be of order $\lambda/{\cal L}$, so that the hypothesis of
locality would be recovered in the eikonal limit $\lambda/{\cal L}\rightarrow 0$.  It is a
general property of the Volterra system (5) that for continuous functions there is a unique
relationship between ${\cal F}_{\alpha\beta}$ and $F_{\mu\nu}$.  The acceleration is usually
assumed to be turned on at some initial time and then turned off after a finite duration of
proper time in order to avoid unphysical situations such as the infinite energy required to
keep a hyperbolic observer of uniform acceleration $g$ in motion for all time.  Once the
acceleration is turned off, the observer measures a constant additional field that is the
residue of past acceleration; in fact, such a constant memory field is always allowed since
Maxwell's equations are linear partial differential equations and any solution is determined up
to a constant field.  For a laboratory device, the residue is canceled once the device is reset.

To determine the kernel $\cal K$, it is natural to assume that $\cal K$ is a convolution-type
kernel depending only upon $\tau-\tau^{\prime}$.  We have seen that it is possible for
$F^{\prime}_{\alpha\beta}$ to become a constant under certain circumstances.  According to the
principle developed in the previous section, the measured field ${\cal F}_{\alpha\beta}$ should
never become a constant for an incident radiation field $F_{\mu\nu}$.  To implement this idea,
we recall that for inertial observers the Doppler effect implies that $\omega^{\prime}=0$ only
when $\omega$ vanishes so that once the radiation field is constant according to one observer,
then it must be constant according to all observers.  Generalizing this circumstance to
arbitrary accelerated observers, we conclude that if ${\cal F}_{\alpha\beta}$ is constant, then
$F_{\mu\nu}$ must be constant.  Following this line of thought, we write equation (2) as
$F^{\prime}=\Lambda F$ and equation (5) as 

\begin{eqnarray} {\cal F}(\tau)=F^{\prime}(\tau)+\int^\tau_{\tau_{0}}{\cal K}(\tau -
\tau^{\prime})F^{\prime}(\tau^{\prime})d\tau^{\prime}\;\;,
\end{eqnarray}

\noindent and we find the following integral equation for the kernel $\cal K$ in terms of
$\Lambda(\tau)$,

\begin{eqnarray}
\Lambda(\tau)+\int^{\tau}_{\tau_{0}}{\cal
K}(\tau-\tau^{\prime})\Lambda(\tau^{\prime})d\tau^{\prime}=\Lambda(\tau_{0})\;\;.
\end{eqnarray}

\noindent This equation can be solved in terms of the {\it {resolvent kernel}} $\cal R$,

\begin{eqnarray}
\Lambda(\tau_{0})+\int^{\tau}_{\tau_0}{\cal R}(\tau -
\tau^{\prime})\Lambda(\tau_0)d\tau^{\prime} = \Lambda(\tau)\;\;,
\end{eqnarray}

\noindent which implies that 

\begin{eqnarray} {\cal
R}(\theta)=\frac{d\Lambda(\tau_{0}+\theta)}{d\theta}\Lambda^{-1}(\tau_0)\;\;.
\end{eqnarray}

\noindent Thus the resolvent kernel is proportional to the acceleration of the observer.  The
kernel $\cal K$ can be expressed in general in terms of an infinite series in the resolvent
kernel
$\cal R$; equivalently, $\cal K$ can be determined via $\cal R$ by means of Laplace
transforms.  If the observer is inertial, ${\cal R}=0$ and hence ${\cal K}=0$ and the standard
theory of Lorentz invariance is recovered.

Our treatment (6) - (9) is valid for any field $F$, though we have considered electromagnetism
for the sake of concreteness.  Moreover, the kernel ${\cal K}$ is in general nonzero except for
constant
$\Lambda$ which is the case for a  scalar (or a pseudoscalar) field.  Thus a scalar field is local
according to this theory.  Hence a fundamental scalar field can stand completely still
with respect to an accelerated observer.  This is contrary to the principle formulated in the previous
section; therefore, a basic scalar field is excluded by the nonlocal theory [26].  It thus
follows from the nonlocal theory that any scalar field found in nature must be a composite.

It is important to subject the nonlocal theory to direct experimental test.  The current status
of this problem is considered in the next section.
\vspace{12pt}

\noindent 4. DISCUSSION
\vspace{10pt} 

In the thought experiment employed in section 1 to illustrate spin-rotation coupling for
radiation received by a uniformly rotating observer, the nonlocal contribution to the amplitude
of the measured radiation constitutes a direct test of the nonlocal theory.  It turns out that
for the experimentally viable case of $\omega\gg\Omega$, for example, there is a relative
increase (decrease) in the measured amplitude by $\Omega/\omega$ as a consequence of nonlocality
for incident RCP (LCP) waves [26].  In the JWKB limit, however, $\Omega/\omega =
\lambdabar/{\cal L}\rightarrow 0$  and the result of the standard theory is recoverd, as expected. 
This effect may be searched for -- in the rotating frame -- in order to test the nonlocal
theory; however, the influence of rotation on the measuring device must then be taken into
account.  The problems associated with the standard electrodynamics of accelerated
{\it {media}} are quite nontrivial.  The assumptions that are usually employed in the
design of electrical equipment have been reviewed by Van Bladel [27].  It therefore appears
that the proposed search for nonlocality of order $\Omega/\omega\sim 10^{-8}$ in the rotating
system would have to involve rather delicate experiments [26].  To circumvent such problems,
Shoemaker [28] has proposed a test of nonlocal electrodynamics in the laboratory (i.e.
inertial) frame.  

In view of the above remarks, let us therefore consider the problem of testing the nonlocal
theory in a different context: instead of an observer in a rotating system, let us imagine an electron
in a Rydberg state of high angular momentum.  In the correspondence limit, the interaction of the
incident radiation field with the electron would be expected to reflect the nonlocal effect under
consideration here.  It is therefore interesting to search for evidence in connection with the
nonlocal theory in the standard quantum treatment of atomic transitions such as the photoelectric
effect.  The polarization dependence of the photoelectric effect has recently received attention in
connection with the angular distribution of the electrons that are ejected as a result of the
interaction of atoms with x-rays from synchrotron light sources [29].  To test the nonlocal theory,
it appears necessary to study the explicit form of the total cross section for the photo-effect in
the case of incident circularly polarized radiation.  In this regard, it is interesting to note that 
the {\it impulse approximation} of quantum scattering theory [30] is physically equivalent to the
hypothesis of locality.  Therefore, it is in general necessary to go beyond the impulse approximation
and include the influence of the Coulomb interaction explicitly.  These issues require further
investigation.
\vspace{12pt}

\noindent ACKNOWLEDGEMENT
\vspace{10pt}

I am grateful to S. Chu and M. Kasevich for a useful discussion.
\vspace{12pt}

\noindent REFERENCES
\vspace{10pt}
\begin{enumerate}
\item See, for example, de Oliveira, C. G., and Tiomno, J. (1962). {\it Nuovo Cimento} {\bf
24}, 672; Mitskievich, N.V. (1969).  {\it Physical Fields in General Relativity Theory} (Nauka,
Moscow); Schmutzer, E. (1973).  {\it Ann. Physik} {\bf 29}, 75; Barker, B.M., and O'Connell,
R.F. (1975).  {\it Phys. Rev. D} {\bf 12}, 329; Schmutzer, E., and Pleba\'{n}ski, J. (1977). 
{\it Fortschr. Phys.} {\bf 25}, 37.
\item Einstein, A. (1955).  {\it The Meaning of Relativity} (Princeton University Press,
Princeton).
\item Mashhoon, B. (1987).  {\it Phys. Lett. A} {\bf 122}, 299; (1992).  {\it ibid}.  {\bf
163}, 7; Beem, J. K., Ehrlich, P.E., and Easley, K. L. (1996).  {\it Global Lorentzian
Geometry}, 2nd Ed. (Dekker, New York), ch. 2.
\item Mashhoon, B. (1973).  {\it Phys. Rev. D} {\bf 7}, 2807; (1974).  {\it ibid} {\bf
10}, 1059; (1974).  {\it Nature} {\bf 250}, 316; (1975).  {\it Phys. Rev. D} {\bf 11}, 2679;
(1993).  {\it Phys. Lett. A} {\bf 173}, 347.
\item Harwit, M., {\it et al.} (1974).  {\it Nature} {\bf 249}, 230; Dennison, B., Dickey,
J., and Jauncey, D. (1976).  {\it Nature} {\bf 263}, 666; Dennison, B., {\it et al.} (1978). 
{\it Nature} {\bf 273}, 33.
\item Damour, T., and Ruffini, R. (1974).  {\it C. R. Acad. Sci. A} {\bf 279}, 971; Feng,
L. L., and Lu, T. (1991).  {\it Class. Quantum Grav.} {\bf 8}, 851.
\item Zannias, T. (1996).  UNAM report; Bogdanov, M. B., Cherepashchuk, A. M., Sazhin, M.
V. (1996).  {\it Astrophys. Space Sci.} {\bf 235}, 219. 
\item Mashhoon, B., Neutze, R., and Hannam, M., to be published. 
\item Mashhoon, B. (1993).  {\it Quantum Gravity and Beyond}, edited by Mansouri, F., and
Scanio, J. (World Scientific, Singapore), p. 257; Chicone, C., Mashhoon, B., and Retzloff, D.
G. (1996).  {\it J. Math. Phys.} {\bf 37}, 3997.
\item Stedman, G. E. (1985).  {\it Contemp. Phys.} {\bf 26}, 311; Anderson, R., Bilger,
H. R., and Stedman, G. E. (1994).  {\it Am. J. Phys.} {\bf 62}, 975.
\item Mashhoon, B. (1988).  {\it Phys. Rev. Lett.} {\bf 61}, 2639; (1992).  {\it ibid.}
{\bf 68}, 3812.
\item Hehl, F. W., and Ni, W. T. (1990).  {\it Phys. Rev. D} {\bf 42}, 2045; Hehl, F. W.,
Lemke, J., and Mielke, E. W. (1991).  {\it Geometry and Theoretical Physics}, edited by
Debrus, J., and Hirshfeld, A. C. (Springer, Berlin), p. 56; Audretsch, J., Hehl, F. W., and
L\"{a}mmerzahl, C. (1992).  {\it Relativistic Gravity Research}, edited by Ehlers, J., and
Sch\"{a}fer, G. (Springer, Berlin), p. 368.
\item Huang, J. (1994).  {\it Ann. Physik} {\bf 3}, 53.
\item Soares, I.D., and Tiomno, J. (1996).  {\it Phys. Rev. D} {\bf 54}, 2808.
\item Singh, P., and Ryder, L. H. (1997).  {\it Class. Quantum Grav.} {\bf 14}, 3513.
\item Ryder, L. H. (1998).  {\it J. Phys. A:  Math. Gen.}, in press.
\item Silverman, M. P. (1991).  {\it Phys. Lett. A} {\bf 152}, 133; (1992).  {\it Nuovo
Cimento D} {\bf 14}, 857.
\item Cai, Y. Q., and Papini, G. (1991).  {\it Phys. Rev. Lett.} {\bf 66}, 1259;
(1992).  {\it ibid.} {\bf 68}, 3811.
\item Bell, J. S., and Leinaas, J. (1983).  {\it Nucl. Phys. B} {\bf 212}, 131; (1987). 
{\it ibid.} {\bf 284}, 488.
\item Cai, Y. Q., Lloyd, D. G., and Papini, G. (1993).  {\it Phys. Lett. A} {\bf 178},
225.
\item Wineland, D. G., {\it et al.} (1991).  {\it Phys. Rev. Lett.} {\bf 67}, 1735;
Venema, B. J., {\it et al.} (1992).  {\it ibid.} {\bf 68}, 135.
\item Mashhoon, B. (1995).  {\it Phys. Lett. A} {\bf 198}, 9.
\item Mashhoon, B. (1988).  {\it Phys. Lett. A} {\bf 126}, 393; (1993).  {\it Found. Phys.
Lett.} {\bf 6}, 545.
\item Minkowski, H. (1952).  {\it The Principle of Relativity}, by Lorentz, H. A.,
Einstein, A., Minkowski, H., and Weyl, H. (Dover, New York), p. 80.
\item Mach, E. (1960).  {\it The Science of Mechanics} (Open Court, La Salle), ch. II,
sec. VI (n. b. part 6).
\item Mashhoon, B. (1993).  {\it Phys. Rev. A} {\bf 47}, 4498; (1994).  {\it Cosmology
and Gravitation}, edited by Novello, M. (Editions Fronti\`{e}res, Gif-sur-Yvette), p. 245.
\item Van Bladel, J. (1976).  {\it Proc. IEEE} {\bf 64}, 301; (1984).  {\it Relativity
and Engineering} (Springer, New York).
\item Shoemaker, G. H. N. (1997).  ``A Test of Covariance and Nonlocal Electrodynamics in
the Laboratory Frame,'' preprint.
\item Peshkin, M. (1998).  ``Photon Beam Polarization and Nondipolar Angular Distributions,''
Argonne preprint; (1996).  In:  {\it Atomic Physics with Hard X-Rays from High Brilliance Synchrotron
Light Sources}, Argonne report ANL/APS/TM-16; (1970).  {\it Advances in Chemical Physics} {\bf 18}, 1.
\item Goldberger, M. L., and Watson, K. M. (1964).  {\it Collision Theory} (Wiley, New
York), p. 683; Gottfried, K. (1966).  {\it Quantum Mechanics} (Benjamin, New York), p. 453.
\end{enumerate}
\end{document}